\documentclass[%
showpacs,
prd,
amsmath,amssymb,
preprint
draft
]{revtex4-1}
\usepackage[pdftex]{graphicx}
\usepackage{bm}
\usepackage{float,color}
\usepackage{comment}
\usepackage{subfig}
\usepackage{bbold}
\usepackage{physics}
\newcommand{\eq}[1]{\begin{align} #1 \end{align}}

\newcommand{\bs}[1]{\boldsymbol{#1}}
\newcommand{\adag}{\hat c^{\dagger}{}_{\!\!\bm{k}}}
\newcommand{\adagg}{\hat c^{\dagger}{}_{\!\!\!-\bm{k}}}
\newcommand{\pa}{\partial}

\begin{document}
\title{The decoherence and interference of cosmological arrows of time for a de Sitter universe with quantum fluctuations}
%
%
\author{Marcello Rotondo}
\email{marcello@gravity.phys.nagoya-u.ac.jp}
\author{Yasusada Nambu}
\email{nambu@gravity.phys.nagoya-u.ac.jp}
\affiliation{Department of Physics, Graduate School of Science, Nagoya 
University, Chikusa, Nagoya 464-8602, Japan}
\date{May 5, 2018}%
\begin{abstract}
We consider the superposition of two semiclassical solutions of the Wheeler-DeWitt equation for a de Sitter universe, describing a quantized scalar vacuum propagating in a universe that is contracting in one case and expanding in the other, each identifying a opposite cosmological arrow of time. We discuss the suppression of the interference terms between the two arrows of time due to environment-induced decoherence caused by modes of the scalar vacuum crossing the Hubble horizon. Furthermore, we quantify the effect of the interference on the expectation value of the observable field mode correlations, with respect to an observer that we identify with the spatial geometry.
\end{abstract}
%
\maketitle
%
\section{Introduction}
Resting on the assumption that the laws of Nature are fundamentally
quantum, quantum cosmology is the attempt to extend the quantum
formalism to the whole universe.  This attempt must naturally
accommodate the properties of the classical universe we observe and it
should as well provide predictions about the conditions of the very
early universe where a drastic divergence from the classical theory
might be expected.  One of the main approaches to quantum gravity and
cosmology, the canonical one, starts with the quantization of the
Hamiltonian formulation of general relativity, resulting in a momentum
constraint enforcing the diffeomorphism invariance of the theory and
a Hamiltonian constraint encoding the dynamics of the geometry
\cite{deWitt}. Initially \cite{DeWitt0}\cite{Rovelli} the quantization
was formulated in the geometrodynamical representation, were the
canonical variables are the spatial metric components of the chosen
foliation and the momenta conjugated to them, which are related to the
extrinsic curvature. Although later introduction of different
variables (Ashtekar variables and finally loop variables) helped to
solve many serious technical problems \cite{Rovelli0} and lead to a
promising loop quantum cosmology, in this work we will adopt the
intuitive picture provided by the geometrodynamical representation.

Notwithstanding the problems of this representation, the
Wheeler-DeWitt equation (the quantized form of the Hamiltonian
constraint) and its solution (a complex functional of the 3-metrics
and matter field configurations spanning the so-called ``superspace'')
can be ``cured'' by considering specific ``minisuperspace'' models,
that is models obtained by dealing only with a finite number of
homogeneous degrees of freedom of the whole theory and dropping all
the others. The most common choice is to consider a closed Friedmann
universe defined simply by a scale factor $a$ and inhabited by a
homogeneous scalar field $\phi$. Then, the wave functional is usually
interpreted as a true ``wave function of the universe'', such as in
the most notable cases of the Hartle-Hawking state \cite{Hawking} or
the tunneling wave function by Vilenkin
\cite{Vilenkin}.

The construction of the Wheeler-DeWitt equation was suggested by a
quantization of the Einstein-Hamilton-Jacobi equation \cite{Peres} of
general relativity in analogy with the construction of the
Schr\"{o}dinger equation from the Hamilton-Jacobi equation of
classical mechanics.  By its own construction, then, in the
semi-classical approximation where the main contribution to the action
comes only from the gravitational degrees of freedom (d.o.f.) and the
non-gravitational d.o.f. can be treated as perturbations, the wave
functional can be factorized by performing an expansion in orders of
$M=1/32\pi G$ \cite{Kiefer1}, in analogy with the Born-Oppenheimer
expansion for electrons bound to massive nuclei.  This results up to
the second order in a WKB-like approximation where the
``semiclassical'' factor depends only on the spatial geometry, with
the components of the spatial metric playing the role of ``slow''
d.o.f.  satisfying the vacuum Einstein field equations, while the
``full quantum'' factor depends also on the ``fast'' perturbative
d.o.f., for which the usual time evolution in the form of
the Tomonaga-Schwinger equation is naturally retrieved.

In the present work, we consider the superposition of two such
solutions in the de Sitter universe, each one defining a opposite
``arrow of time'' \cite{Zeh} \cite{Zurek} corresponding to a
contracting and a expanding universe. We discuss their quantum
decoherence ( i.e. the suppression of the interference terms of their
superposition), when they are coupled to quantum fluctuations of a
scalar vacuum. In this case, the decoherence follows from the fact
that the modes of these fluctuations that are beyond the Hubble scale
cannot be observed. We can treat therefore these modes as a external
environment and the modes inside the horizon as a probe of the spatial
geometry.  Decoherence of the wave function of the universe, also in
relation with the arrow of time, has already been studied since works
such as \cite{Zeh0}--\cite{Kiefer2}. The contribution of the present
work is twofold. First, we do not drop all the infinite quantum
d.o.f. of the scalar vacuum and consider its second quantization,
which was not done in previous works. In this case, the coupling to
geometry leads to gravitational particle production
\cite{Davies}. Secondly, we quantify the effect of the interference of
cosmological arrows of time on the expectation value of the observable
field mode correlations, with respect to an observer that we identify
with the spatial geometry.
%
\section{Semiclassical approximation of geometrodynamics in de Sitter universe} \label{1}

We will briefly review the semiclassical approximation of
geometrodynamics. As introduced before, in this formulation of quantum
cosmology the state of the universe is described by a wave functional
$\Psi = \Psi[\bs{\sigma}, \phi]$ defined on superspace of 3-metrics
$\bs{\sigma}(x)$ and matter fields configurations $\phi(x)$ on the
associated 3-surfaces. This representation of the wave functional is a
solution of the Wheeler-DeWitt equation in its geometrodynamical form
(integration over space is implied as usual)
\eq{
	\left( - \frac{\hbar^2}{2M} \nabla^A\nabla_A - 2
          M \sqrt{\sigma} \left({}^{(3)} R - 2\Lambda \right)  +
          \mathcal{H}_\phi \right) \Psi = 0\label{eq:WDW} \, , \qquad
        \nabla^A \nabla_A := G^{AB} \frac{\delta}{\delta
          \sigma^{A}}\frac{\delta}{\delta \sigma^{B}}  \, . 
}
Here
$G^{AB} \equiv G^{ijkl} = \sqrt{\sigma} \left( \sigma^{ik}\sigma^{jl} + \sigma^{il}
  \sigma^{jk} - 2\sigma^{ij} \sigma^{kl} \right)/2 $
is the de Witt supermetric (latin capital letters representing pairs
of spatial indexes and $\sigma$ being the determinant of the spatial
metric), ${}^{(3)}R$ is the Ricci scalar of the spatial metric,
$\Lambda$ is the cosmological constant and $M=1/32\pi G = (M_P/2)^2$
with $M_P$ being the reduced Planck mass (in the following we will set
$\hbar = c = 1$), while
$\mathcal{H}_\phi=\mathcal{H}_\phi[\bs{\sigma}, \phi]$ is the
Hamiltonian density for the matter fields (in the following we will
have only one matter scalar field for simplicity). While technical
features such as the absence of a clear scalar product for the
wavefunctional and the divergences that may appear while handling the
functional derivatives make the Wheeler-DeWitt not well-behaved in its
general form (see e.g. \cite{Woodard} for the factor-ordering problem for the product of local operators acting at the same space point), the equation is well-defined within proper cosmological
minisuperspace models such as the one we consider.

In the oscillatory region of the wave function, associated to
classically allowed solutions, one may consider the ansatz
\cite{Halliwell}
\begin{equation}
 	\Psi[\bs{\sigma}, \phi] = C[\bs{\sigma}] \exp\left( i\,
          M\, S[\bs{\sigma}] \right) \chi[\bs{\sigma},
        \phi] \label{eq:Psi} 
\end{equation}
where $C[\bs{\sigma}]$ is a slowly varying prefactor and
$\chi[\bs{\sigma}, \phi]$ represents small perturbations of order
$(M){}^0$ to the action $M\,S[\bs{\sigma}]$. For this
ansatz, at order $(M)^{-1}$ the Wheeler-DeWitt equation gives the
Einstein-Hamilton-Jacobi equation for the action $S[\bs{\sigma}]$
\eq{
\frac{1}{2}  \nabla^A S \nabla_A S + V = 0, \qquad V = - 2\sqrt{\sigma} \left({}^{(3)}R - 2\Lambda \right), \label{eq:M1}
}
as well as the current conservation equation for the density $C^2$
\eq{
	\nabla^A S \nabla_A C + \frac{C}{2} \nabla^A \nabla_A S = 0 \label{eq:cc} \,
}
with geometrodynamical momentum 
\eq{
	\pi^{A} \equiv M\nabla^A S  \label{eq:g_mom} \, 
}
conjugate to $\sigma^A$ and proportional to the extrinsic curvature of
the geometry.  At order unity $(M)^0$, the Wheeler-DeWitt equation
gives (see e.g.  \cite{Lapchinsky} for a formal derivation)
\eq{
	i \nabla^A S \nabla_A \chi - \mathcal{H}_{\phi} \chi = 0 \label{eq:ST0} \, 
}
where integration over space is still implied. It is convenient to
introduce the vector tangent to the set of solutions determined by
\eqref{eq:g_mom},
\eq{
	\frac{\delta}{\delta t} := \nabla^A S \nabla_A \label{eq:WKB_t}
}
where the affine parameter along the trajectories,
$t[\sigma_{ij}(x)]$,
is found to correspond to usual time and usually referred to as
``WKB'' or ``bubble'' time $t$. Then \eqref{eq:ST0} corresponds to the
Tomonaga-Schwinger equation for the matter field, $\mathcal{H}_\phi$
being the Hamiltonian density of interaction between geometrodynamical
and matter degrees of freedom.

In the present case, we are interested in a spatially flat FLRW universe.
%
It is convenient to write the 3-metric in the form
$\sigma_{ij}=\sigma^\frac{1}{3} \tilde\sigma_{ij}$, which gives the EHJ
equation \eqref{eq:M1} in the form \cite{Kiefer1}
\eq{
	-\frac{3}{16} \sqrt{\sigma} \left( \frac{\delta S}{\delta \sqrt{\sigma}}\right)^2 + \frac{1}{2} \frac{\tilde{\sigma}_{ik}\tilde{\sigma}_{jl}}{\sqrt{\sigma}} \frac{\delta S}{\delta{\tilde{\sigma}_{ij}}}\frac{\delta S}{\delta{\tilde{\sigma}_{kl}}}-2\sqrt{\sigma} ({}^{(3)}R - 2 \Lambda) = 0 \, . \label{eq:HJEdS}
}
For the de Sitter universe with spatially flat slicing, we have
${}^{(3)}R = 0$ and $S = S(\sqrt{\sigma})$ if we neglect the
  contribution of the tensor mode $\tilde\sigma_{jk}$, which gives
\eq{
	S = S_{\pm} = \pm 8 \sqrt{\frac{\Lambda \sigma}{3}} \label{eq:S0} \, ,
}
where we have fixed the spatial volume $V_0=\int d^3x$ to unity and
the sign ambiguity is due to the invariance under time inversion of
the Einstein-Hamilton-Jacobi equation, the minus sign corresponding to
the arrow of time of a expanding universe and the plus sign
corresponding to the arrow of time of a contracting one. Inserted in
\eqref{eq:cc}, \eqref{eq:S0} gives a constant prefactor $C$ for the de
Sitter universe and the WKB time it defines through \eqref{eq:WKB_t}
for an expanding universe ($S_{-}$) is
\eq{
	\frac{\delta}{\delta t} = - \frac{3}{8} \sqrt{\sigma} \frac{\delta S_\pm}{\delta \sqrt{\sigma}} \frac{\delta}{\delta \sqrt{\sigma}} = \mp \sqrt{3 \sigma \Lambda} \frac{\delta}{\delta \sqrt{\sigma}} \, . \label{eq:WKB}
}
One can check that \eqref{eq:HJEdS} is equivalent to the vacuum
Friedman equation and that the WKB time \eqref{eq:WKB} is the Friedman
time providing the usual expression for the expansion scale factor
$a(t)=\exp(H t)=\sigma^{1/6}$, with Hubble constant given by
$H=\sqrt{\Lambda/3}$. The chosen metric is
\begin{equation}
ds^2 =  -dt^2 + a^2(t)d\bm{x}^2. \label{eq:FLRW}
\end{equation}
which, introducing the conformal time $\eta \in (-\infty,0) $ as
$d\eta=dt/a$, can also be written in the conformal form
\begin{equation}
  ds^2=a^2(\eta)(-d\eta^2+d\bs{x}^2),\quad a(\eta)=-1/H\eta \, .
\end{equation}
\section{Decoherence of different arrows of time} \label{2}
We will consider the interference between the two different WKB
branches associated to opposite sign of actions, taking a real
solution of the vacuum Wheeler-DeWitt equation in the oscillatory
regime
\eq{
	\Psi^{(0)}(a) \propto  e^{ iMS_+(a)} + e^{ i M S_-(a)}  \, , \label{eq:psi0}
}
which corresponds to a superposition of a contracting and an expanding
universe. The generalization of the following results to other choices
is straightforward.  For simplicity of notation we make explicit only
the dependency of the action $S$ on the conformal factor although it
is a function of its time derivative as well. We consider then a
perturbation of the action represented by the wave functional of some
non-gravitational field modes
$\chi(\varphi_{\bs{k}},\varphi_{-\bs{k}})$, the precise form of which
we will derive later. The total wave functional will become
\begin{align}
	\Psi^{(0)}(a) \rightarrow 
	\Psi(a;\varphi_{\bs{k}},\varphi_{-\bs{k}}) =
	\frac{1}{\sqrt{2}} \left( e^{ i M
		S_+(a)}\chi^*(a;\varphi_{\bs{k}},\varphi_{-\bs{k}}) + e^{ i
		M
		S_-(a)}\chi(a;\varphi_{\bs{k}},\varphi_{-\bs{k}})
	\right) \label{eq:psi} 
\end{align}
since in the contracting and expanding branches the wave functionals of
the perturbations are solutions of the Tomonaga-Schwinger equation and
its conjugate, respectively. These perturbations can be used in a
sense as a probe of the gravitational field to which they are coupled.

In addition to interference, we will take into account a mechanism for
the decoherence of the two arrows of time. A classic example of
quantum decoherence is provided in the context of the widely discussed
problem of measurement in quantum mechanics. We do not intend to give
here a full review nor to focus on the measurement problem, for which
we refer to rich classic literature \cite{Joos0}. In the ideal von
Neumann measurement scheme, a given system $\mathcal{S}$ represented
in the basis $\{\ket{s_i}\}$ of the Hilbert space
$\mathcal{H}_\mathcal{S}$ interacts with a measurement apparatus
$\mathcal{A}$ analogously represented by the basis of pointer vectors
$\{\ket{a_i}\} \in \mathcal{H}_\mathcal{A}$ each corresponding to a
outcome reading of the apparatus associated to the state $\ket{s_i}$
of $\mathcal{S}$. While the system and apparatus are initially
uncoupled, with the latter in a certain ``ready'' state $\ket{R}$,
after some time $t$ (short enough to allow to neglect here the
self-evolution of the state) they will evolve into the
``pre-measurement'' unseparable state
\eq{
	\rho_{\mathcal{S}\mathcal{A}} = \sum_{i,j} c_i\, c_j^* \ket{s_i}\bra{s_j}  \otimes \ket{R}\bra{R} \longrightarrow  \sum_{i,j} c_i\, c_j^* \ket{s_i}\bra{s_j} \otimes \ket{a_i}\bra{a_j}\, .
}
What is usually identified as the measurement problem is the so called
``problem of definite outcomes'', which questions why the apparatus is
in practice found in a specific reading $\ket{a_i}$ among all possible
ones. (An other aspect of the measurement problem consists in the
ambiguity in the freedom we have to chose the representations
$\ket{a_i}$ and $\ket{s_i}$, which also finds its place in the
treatment.)

While decoherence does not explain how the apparatus collapses on a
specific reading, it does provide an explanation for the transition
from quantum amplitudes to classical probabilities for its outcomes by
adding a environment $\mathcal{E}$ represented in a nearly orthonormal
basis $\{\ket{e_i}\}$. By doing so, the evolution of the initial
tripartite state will be
\begin{align}
	\rho_{\mathcal{S}\mathcal{A}\mathcal{E}} 
	= \sum_{i,j} c_i\,
                                                   c_j^*
                                                   \ket{s_i}\bra{s_j}
                                                   \otimes
                                                   \ket{R}\bra{R}
                                                   \otimes
                                                   \ket{E}\bra{E} 
	 \longrightarrow \sum_{i,j} c_i\, c_j^* \ket{s_i}\bra{s_j} \otimes \ket{a_i}\bra{a_j} \otimes \ket{e_i} \bra{e_j} \, .
\end{align}
The important observation at this point is that the environmental
degrees of freedom are not accessible to the observer of the system,
so that the system will be effectively described by the reduced
density matrix
\eq{
	\rho_{\mathcal{S}\mathcal{A}}' = \mathrm{Tr}_{\mathcal{E}}\left[\rho_{\mathcal{S}\mathcal{A}\mathcal{E}}\right] = \sum_{i,j} c_i\, c_j^* \ket{s_i}\bra{s_j} \otimes \ket{a_i}\bra{a_j} \braket{e_i}{e_j} \label{eq:dec}
}
obtained by tracing out the environmental degrees of freedom. The
decay of the off-diagonal elements associated to interference is due
to the vanishing of $\braket{e_i}{e_j}$.

The same principle of environment-induced decoherence can be applied
to the arrows of time, where the measured system can be identified
with the de Sitter geometry while the measuring apparatus and the
environment can be identified respectively with field modes inside and
outside the Hubble horizon, the latter being inaccessible to the
observer.  We will consider as a simple example the perturbations
described by a free massless scalar perturbations $\phi$ of a vacuum
minimally coupled to the metric of the pure de Sitter universe. This
could represent scalar perturbations of the metric or perturbations of
a non-dynamical inflaton field, although we will continue to refer to
it as matter or non-gravitational d.o.f. . Rescaling this physical
field as $\varphi=a\,\phi$, the perturbation of the geometrical action
$M\,S$ is determined by the Lagrangian
\begin{equation}
L=\frac{1}{2} \int d^3x \left( \left(\varphi' - \frac{a'}{a}
\varphi \right)^2 - (\partial_i \varphi)^2 \right)
\end{equation}
(where $'=\partial/\partial\eta$), which gives equations of motion of the form
\begin{equation}
\varphi'' - \left(\pa_i^2+ \frac{a''}{a} \right) \varphi= 0 \, .
\label{eq:eom} 
\end{equation}
We introduce the Fourier mode decompositions of the field and its
conjugate momentum as
\begin{align}
\varphi(\eta,{\bm{x}})= 
\int \frac{d^3k}{(2\pi)^{3/2}}\, \varphi_{\bm{k}}(\eta)\,
e^{i\bm{k}\cdot\bm{x}},\quad
p(\eta,{\bm{x}})  = 
\int \frac{d^3k}{(2\pi)^{3/2}}\, p_{\bm{k}}(\eta)\, e^{i \bm{k} \cdot
	\bm{x}},
\end{align}
where
$p_{\bm{k}}= \varphi_{\bm{k}}'^*-H_c\varphi_{\bm{k}}^*$, $H_c = a'/a =
aH$ being the conformal Hubble constant. Following the usual
quantization procedure, we express the associated operators in terms
of the time-dependent creation and annihilation operators
$\hat{c}_{\bm{k}}(\eta)^{\dag}$ and $\hat{c}_{\bm{k}}(\eta)$
\begin{align}
\hat{\varphi}_{\bm{k}}  = \frac{1}{\sqrt{2k}} \left(
\hat{c}_{\bm{k}}
+ \adagg \right),\quad
\hat{p}_{\bm{k}} = -i\sqrt{\frac{k}{2}}\, 
\left( \hat{c}_{\bm{k}} - \adagg \right),
\label{eq:cre}
\end{align}
so that the Hamiltonian $\hat{H}_\phi = \int dx^3
\,\hat{\mathcal{H}}_{\phi}$ can be expressed as 
\begin{align}
\hat {H}_{\phi} (\eta)= \int_{(\mathbb{R}^{+})^3} \!\! d^3k\left[\frac{k}{2}\,(\adag\,\hat c_{\bm{k}}
+\hat c_{\bm{k}}\,\adag) + 
i H_c\,(\adag\,\adagg-\hat c_{\bm{k}}\,\hat c_{-\bm{k}})\right]\, , \label{eq:H}
\end{align}
where $k = |\bs{k}|$.The first term of \eqref{eq:H} corresponds to the usual free time-dependent Hamiltonian while the second term provides the coupling
of the gravitational and non-gravitational d.o.f. and determines the
production of ``particles'' through squeezing of the initial vacuum
state due to the expansion (squeezing along the field quadrature) or
contraction (squeezing along the momentum quadrature) of the
universe. The strength of the coupling is determined by the conformal
Hubble
constant $H_c$. 

The Heisenberg equations of motion for the operators $\hat{c}_{k}$ and
$\hat{c}_{k}^{\dag}$ for the two modes are
\begin{align}
	\hat{c}_{\bm{k}}{}' = i \left[ \hat {\mathcal{H}}_{\phi} , \hat{c}_{\bm{k}}\right] 
	= - i k\, \hat{c}_{\bm{k}} + \frac{a'}{a} \adagg \, , \qquad
	\adagg{}' = i \left[ \hat {\mathcal{H}}_{\phi} ,
	\adag\right] =
	i k\,\adagg + \frac{a'}{a} \hat{c}_{\bm{k}} \, .
	\label{eq:Heis_op}
\end{align}
The operators at conformal time $\eta_0$ are related to those at a later time $\eta>\eta_0$ by a Bogolyubov transformation
\begin{equation}
\binom{\hat{c}_{\bm{k}}(\eta)}{\adagg(\eta)}
= \binom{\alpha_k(\eta) \ \ \ \beta_k(\eta)}{\beta_k^{*}(\eta) \
	\ \ \alpha^*_k(\eta)}
\binom{\hat{c}_{\bm{k}}(\eta_0)}{\adagg(\eta_0)}
\quad , 
\label{eq:bog_matrix}
\end{equation}
where $\alpha_k$ and $\beta^{*}_k$ are the Bogolyubov coefficients
satisfying $|\alpha_k|^2 - |\beta_k|^2= 1$, with the initial
conditions $\alpha_k(\eta_0) = 1 $ and $\beta_k(\eta_0) = 0$ (here and in the following we use the notation $k = |\bf{k}|$). The time
evolution of these coefficients is obtained straightforwardly from
\eqref{eq:Heis_op}
\begin{align}
	\alpha_k{}' = - i k\,\alpha_k + \frac{a'}{a} \beta^*_k,\quad
	\beta_k{}' = - i k\,\beta_k + \frac{a'}{a} \alpha_k^{*} \, ,
	\label{eq:bogeq}
\end{align}
and in the case of de Sitter universe with vacuum fixed at
$\eta_0\to-\infty$ they are given by 
\eq{
	\alpha_k = \left(1 - \frac{i}{2k\eta}\right) e^{-ik\eta} \, , \quad \beta_k = \frac{i}{2k\eta} e^{ik\eta} \, .
}
In general, the vacuum state can be defined as the eigenstate of the
annihilation operator at time $\eta_0$ 
\begin{equation}
\hat{c}_{\bm{k}}(\eta_0)|0_{\bs{k}}\rangle_{\text{in}} = 0 \, .
\label{eq:vacuum0}
\end{equation}
Introducing the Schr{\"o}dinger picture of the state at $\eta$, the
transformation \eqref{eq:bogeq} gives the vacuum condition as 
\begin{align}
	\label{eq:vacuum}
	\left( \alpha^{*}_k(\eta) \hat{c}_{\bm{k}} 
	- \beta_k(\eta) \adagg \right) \ket{0_{\bs{k}}}_{\text{out}} = 0
\end{align}
In the basis that diagonalizes $\{ \hat{\varphi}_{\bs{k}},
\hat{\varphi}_{-\bs{k}} \}$, the vacuum condition \eqref{eq:vacuum}
together with \eqref{eq:cre} provides a Gaussian wave function for
each mode of the out-state 
\begin{align}
	&\chi_k(a;{\varphi}_{\bs{k}}, {\varphi}_{-\bs{k}}) =
        \braket{{\varphi}_{\bs{k}},{\varphi}_{-\bs{k}}}{0_{\bs{k}}}_{\text{out}} =
        \left(\frac{2}{\pi} \mathrm{Re}(\xi_k) \right)^{\frac{1}{2}} \exp
        \left(- \xi_k\, {\varphi}_{\bs{k}}\,{\varphi}_{-\bs{k}}
          \right),\\
         & \qquad\qquad \xi_k = \frac{1}{k} \frac{\alpha_k^* -
          \beta_k}{\alpha_k^* + \beta_k } = \frac{k}{1 - i
           H_c/k}\, . \notag 
\end{align}
Notice that for $H\to0$ these will reduce to the Gaussian vacuum
fluctuations of Minkowski spacetime.

The ket vector describing the universe will therefore be
\eq{
	\ket{\Psi} &= \frac{1}{\sqrt{2}} \overbrace{\ket{S_{+}}} \otimes \overbrace{\bigotimes_{\nu = 0}^{a} \ket{0^*_{\bs{k}}}_{\text{out}}} \otimes \overbrace{ \bigotimes_{\nu = a}^\infty \ket{0^*_{\bs{k}}}_{\text{out}}} \notag \\
	& + \frac{1}{\sqrt{2}} \underbrace{\ket{S_{-}}}_{\text{system}} \otimes \underbrace{\bigotimes_{\nu = 0}^{a} \ket{0_{\bs{k}}}_{\text{out}}}_{\text{environment}} \otimes \underbrace{\bigotimes_{\nu = a}^\infty \ket{0_{\bs{k}}}_{\text{out}}}_{\text{apparatus}}
}
where we have used the notation $\braket{\bs\sigma}{S_{\pm}} \simeq
e^{i MS_{\pm}(a)}$, we have defined the dimensionless index $\nu =
k / H$ and $\bigotimes_{\nu = \nu_1}^{\nu_2}$ stands for a tensor product over $\forall \bf{k}$ such that $|{\bf{k}}| = k \in (H\nu_1,H\nu_2)$.

\begin{figure}[H]
	\centering
	\includegraphics[width=0.75\linewidth]{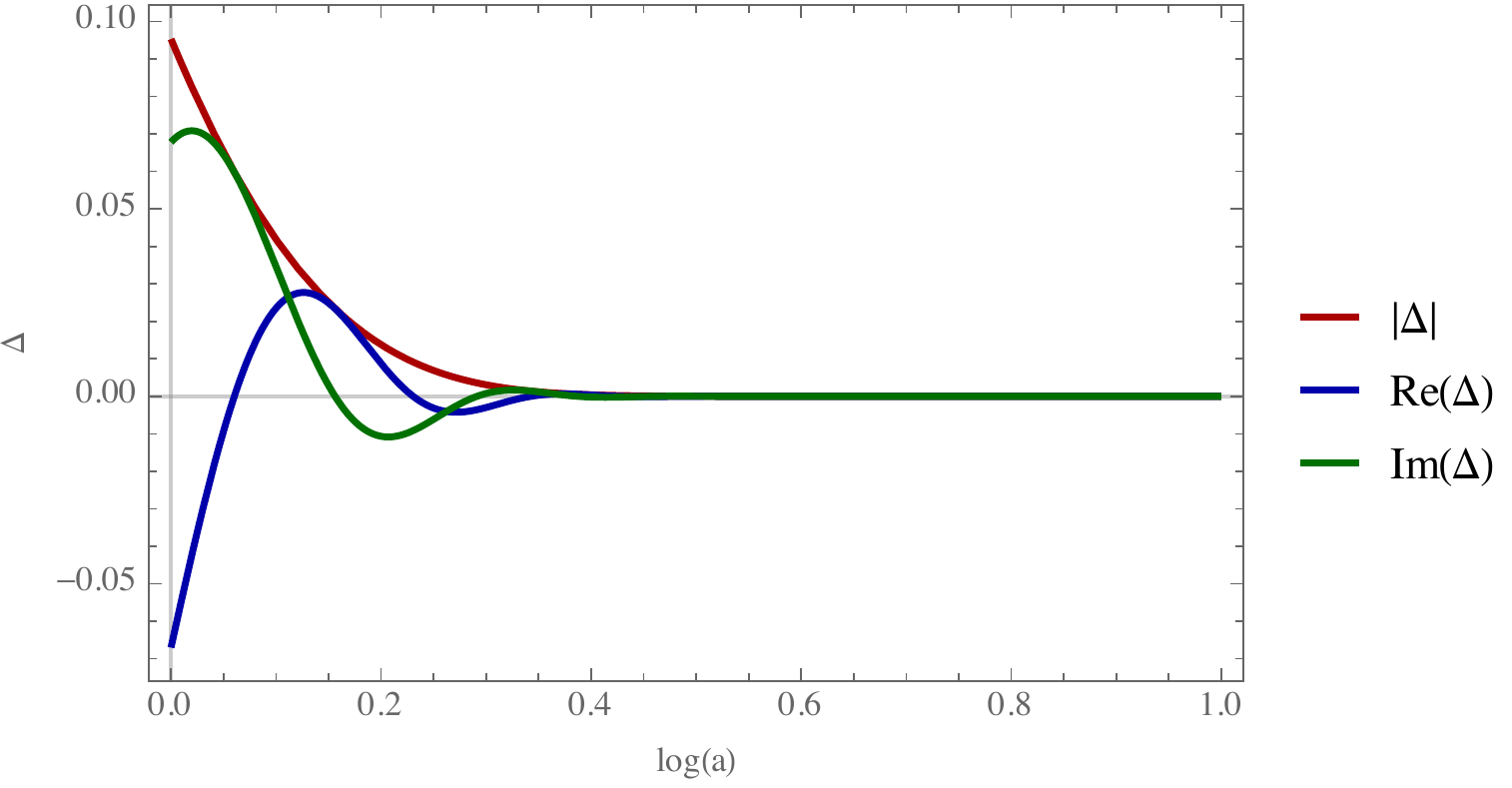} \centering
	\caption{Plot of the damping factor $\Delta$ as a function of the e-folding $\log(a)$.}
	\label{fig:f0}
\end{figure}

For a given $a$, the reduced density matrix $\rho_R$ obtained by
tracing out the environment (i.e. field modes with $k \leq H_c$) will
be
\eq{
	\rho_R & = \frac{1}{2} \left( \ketbra{S_{+}}{S_{+}} \otimes \bigotimes_{\nu = a}^\infty \ket{0^*_{\bs{k}}}{}_{\text{out}}\bra{0^*_{\bs{k}}} + \ketbra{S_{-}}{S_{-}} \otimes \bigotimes_{\nu = a}^\infty \ket{0_{\bs{k}}}{}_{\text{out}}\bra{0_{\bs{k}}} \right) \label{eq:r1} \\
	& + \frac{1}{2} \left( \ketbra{S_{+}}{S_{-}} \otimes \bigotimes_{\nu = a}^\infty \ket{0^*_{\bs{k}}}{}_{\text{out}}\bra{0_{\bs{k}}} \times \Delta^* + \ketbra{S_{-}}{S_{+}} \otimes \bigotimes_{\nu = a}^\infty \ket{0_{\bs{k}}}{}_{\text{out}}\bra{0^*_{\bs{k}}} \times \Delta \right) \label{eq:r2} \,  
}
where $\Delta$ in the last two terms is the damping factor (analogous
to $\braket{e_i}{e_j}$ in \eqref{eq:dec}) that determines the
decoherence of the arrows of time as a function of the scale factor
$a$
	\eq{ 
	\Delta(a) & = \prod_{0}^{a} \left({}_\text{out} \braket{0^*_{\bs{k}}}{0_{\bs{k}}}{}_\text{out}\right)^{d^3\nu} = \prod_{0}^{a} \left( \frac{\mathrm{Re}(\xi_k(a))}{\xi_k(a)} \right)^{d^3\nu} = \prod_{0}^{a} \left( \frac{1}{1 + i a/\nu}\right)^{d^3\nu} = e^{ i (n-2\pi) a^3 - n a^3},
}
where $n=\pi (4 + \ln(4) - \pi)/3 \approx 2.35$ and we have used
$\prod_{a}^{b} f(x)^{dx} = \exp \left( \log \left(
    \prod_{a}^{b} f(x)^{dx} \right) \right) = \exp \left(
  \int_a^b \log f(x) dx \right) $
for the integral product over the dimensionless index $x$. Notice also that using the spherical symmetry of the problem, we have $\int_{(\mathbb{R}^{+})^3} d^3\nu = 4\pi \int_{\mathbb{R}^{+}} \nu^2 d\nu$. The plot
for $\Delta(\log(a))$ is shown in Fig.\ref{fig:f0}. The associated
decoherence time (time corresponding to a damping factor of $1/e$) is
of the order of one tenth of the Hubble time
\eq{
	t_d = \frac{1}{3H}\log\left[(1+n)/n\right] \simeq 0.12 / H \, .
}
%

\section{Observation}
\begin{figure}[H]
	\centering
	\includegraphics[width=0.75\linewidth]{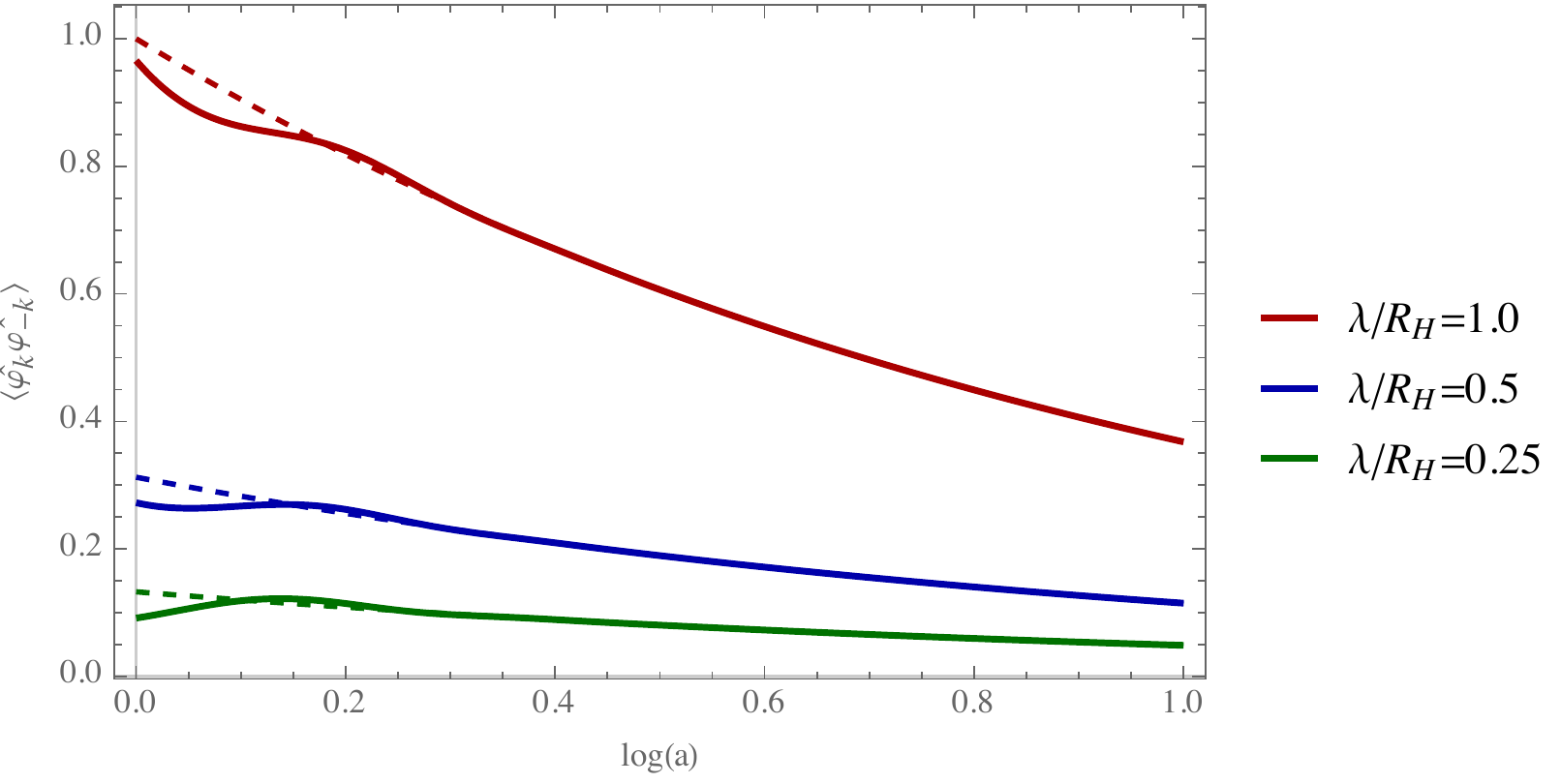} \centering
	\caption{Plot of the field mode correlations for different values of length scales (in units of Hubble radius) as a function of the e-folding (i.e. time in units of Hubble time), superposed to their usual value when interference is not taken into account.}
	\label{fig:f1}
\end{figure}

We consider now the observable effect of the interference of the two
arrows of time in the field mode correlations, which would lead to
observable results in the power spectrum. We will focus here on a
specific mode $k$ and drop all other modes, that are now unobserved
but in principle observable.  Besides the problem of measurement, what
becomes particularly relevant in quantum cosmology is the problem of
the observer: since quantum cosmology is concerned with the total
state of the universe, the question arises whether the observer should
be included in this state and, if so, what the meaning of the wave
function of the universe is.  Discussion of this fundamental issue in
transposing the tools and concepts of quantum mechanics to the study
of the whole universe goes beyond the purpose of the present work, but
one comment is of order if we want to discuss the possible effect of
the interference. In quantum mechanics one usually implicitly assumes
the presence of a \emph{classical} and \emph{external} observer behind
the scenes, separated from the observed system as well as from the
measuring apparatus and/or the environment. This observer does not
``enter the equations'' explicitly. Its implied existence may be seen
in the derivation of the Tomonaga-Schwinger equation \eqref{eq:ST0},
where it can be identified with the 3-geometry and survives only in
the form of the affine (time) parameter along the local worldline of
the ``laboratory''. In this sense, one may say that the observer is
identified with the laboratory's spatial geometry and clock. In the
ansatz \eqref{eq:Psi}, such observers is specified by a single WKB
branch, but in the general solution the observer will be in a
superposition of branches. Unlike the one of non-relativistic quantum
mechanics, this observer is in principle \emph{quantum} and is
\emph{internal} to the wave functional $\Psi$. The semiclassical
approximation then may be seen as the analogous of a large mass limit
of the observer, which is not affected by the observed system but can
still be found in a quantum superposition of two clocks, each ticking
in a different time direction.\\
While we discuss observation from this tentative point of view, the
interpretation of the wave functional of the universe and its observer
deserves to be considered more accurately, possibly along the lines of
a suitable formulation of quantum reference frames
(\cite{Susskind1}--\cite{Aharonov} or \cite{Angelo1}--\cite{Angelo3}
for recent works) or other relational approaches to quantum
observations. Also, works on the concept of ``evolution without
evolution'' such as outlined in \cite{Page} and \cite{Lloyd} may be of
help in clarifying the issue with reference to the emergence of time.
As far as we are concerned here, with what seems a natural extension
from the non-relativistic case to the cosmological case, we
identify the observer with what we previously referred to as the
``system'' and take it to be in the superposition \eqref{eq:psi0}. The
state $\hat{\rho}_{R(\varphi)}$ of the observable matter field for
such observer, up to a proper normalization constant $\mathcal{N}$, is
obtained by projecting $\rho_R$ over such superposition, i.e.
\begin{align}
	\hat{\rho}_{R(\varphi)} \propto & \left( \bra{S_{+}} + \bra{S_{-}} \right)\hat{\rho}_R \left( \ket{S_{+}} + \ket{S_{-}} \right) \notag \\
	 \propto  & \left( \bigotimes_{\nu = a}^\infty
                    \ket{0^*_{\bs{k}}}{}_{\text{out}}\bra{0^*_{\bs{k}}} +
                    \bigotimes_{\nu = a}^\infty
                    \ket{0_{\bs{k}}}{}_{\text{out}}\bra{0_{\bs{k}}}
                    \right) \label{eq:r1b}\\
	+ & \left( \bigotimes_{\nu = a}^\infty \ket{0^*_{\bs{k}}}{}_{\text{out}}\bra{0_{\bs{k}}} \times \Delta^* +  \bigotimes_{\nu = a}^\infty \ket{0_{\bs{k}}}{}_{\text{out}}\bra{0^*_{\bs{k}}} \times \Delta \right)
 \label{eq:r2b}\, .
\end{align}
The expectation value for the correlations of a given field mode
$k \geq H_c$ is then given by two contribution,
$ \expval{\hat\varphi_{\bs{k}}\hat\varphi_{-\bs{k}}} =
\expval{\hat\varphi_{\bs{k}}\hat\varphi_{-\bs{k}}}_{\text{no}} +
\expval{\hat\varphi_{\bs{k}}\hat\varphi_{-\bs{k}}}_{\text{int}} $,
where
\begin{equation}
	\expval{\hat\varphi_{\bs{k}}\hat\varphi_{-\bs{k}}}_{\text{no}} \propto \mathrm{Tr}\left[ \eqref{eq:r1b}_k \hat\varphi_{\bs{k}}\hat\varphi_{-\bs{k}} \right] =  \frac{1}{\mathrm{Re}\left[\xi_k \right]} = \frac{1}{k} \left( 1 + \frac{H_c^2}{k^2} \right)
\end{equation}
is the usual contribution from the terms that survive decoherence, while
\begin{equation}
  \expval{\hat\varphi_{\bs{k}}\hat\varphi_{-\bs{k}}}_{\text{int}}
  \propto  \mathrm{Tr}\left[ \eqref{eq:r2b}_k \hat\varphi_{\bs{k}}\hat\varphi_{-\bs{k}} \right] = \mathrm{Re}\left[ \Delta \frac{ \mathrm{Re}\left[ \xi_k \right]}{{\xi_k^*}^2} \right]  = \frac{1}{k}\mathrm{Re}\left[ \Delta \frac{1+iH_c/k}{1-iH_c/k} \right]  \,
\end{equation}
is the contribution coming from the interference of the arrows of
time.  The normalization constant $\mathcal{N}$ for the diagonal
elements of $\hat{\rho}_{R(\varphi)}$ associated to a given value of
scale factor is
\begin{equation}
	\mathcal{N}(a)^{-1} = 2 \left( 1 + \mathrm{Re}\left[\xi_k(a)\right] \mathrm{Re}\left[\frac{\Delta(a)}{\xi_k(a)}\right] \right)
\end{equation}
which gives normalized correlations (Fig.\ref{fig:f1})
\begin{equation}
 \expval{\hat\varphi_{\bs{k}}\hat\varphi_{-\bs{k}}} = \frac{1}{2k}
  \frac{\left(1 + (H_c/k)^2\right)^2 + \mathrm{Re}\left[ \Delta
      (1+iH_c/k) \right]}{1 +(H_c/k)^2+
    \mathrm{Re}\left[\Delta\left(1-iH_c/k\right)\right]
    } .
\end{equation} 
Using $\varphi_{-\bs{k}}(\eta) = \varphi_{\bs{k}}^*(\eta)$ and
therefore
$\varphi_{+\bs{k}}(\eta) \varphi_{-\bs{k}}(\eta) =
|\varphi_{\bs{k}}(\eta)|^2$,
the probability distribution for the outcome of measurement of the
power spectrum $|\varphi_{\bs{k}}|^2 \in \{0,\infty\}$ is given simply
by
\begin{equation}
	P(|\varphi_{\bs{k}}|^2) \propto \vert\chi_k(a;|{\varphi}_{\bs{k}}\vert^2) \vert^2 + \mathrm{Re} \left[ \chi^2_k(a;|{\varphi}_{\bs{k}}\vert^2) \Delta \right]  \label{eq:ps}\, ,
\end{equation}
which describes how the interference between arrows of time is
expected to affect the usual power spectrum outcome distribution (the
first term in \eqref{eq:ps}) at very early times (that is, early
compared to the Hubble time).
%
\section{Conclusions} \label{3}
We have discussed the interference of two different orientations for
the cosmological arrow of time identified in the expanding and
contracting modes of a de Sitter universe in the semiclassical
approximation of the Wheeler-DeWitt equation, as well as their
decoherence due to quantum fluctuations of a massless scalar vacuum
subject to gravitational particle production. In principle, the
interference of the expanding and contracting modes could be observed
in the field mode correlations and power spectrum at very early times
before the decoherence becomes strong.
Whether this simple model can be applied to the inflationary de Sitter
phase of our universe or other scenarios where the interference of
geometries comes into play is left as a question for future
inquiries. We have provided a tentative reason for identifying the
observer with the spatial geometry and therefore the decoherence
refers to the decoherence of the observer itself, which is coupled to
the scalar vacuum. In any case, besides the details of the decoherence
mechanism, a better understanding of the role of the observer in
quantum cosmology in future works is auspicated.

\acknowledgments{Y. N. was supported in part by JSPS KAKENHI Grant No. 15K05073. M. R. gratefully acknowledges support from the Ministry of Education, Culture, Sports, Science and Technology (MEXT) of Japan.}

\end{document}